\documentstyle[12pt]{article}

\def\trh{T_{\rm RH}}

\def\GeV{\,{\rm GeV}}
\def\TeV{\,{\rm TeV}}

\def\MeV{\,{\rm MeV}}
\def\sec{\,{\rm sec}}

\def\Mpc{\,{\rm Mpc}}

\def\cmm2{{\,\rm cm^{-2}}}
\def\cm2{{\,{\rm cm}^2}}
\def\cmm3{{\,{\rm cm}^{-3}}}
\def\gcmm3{{\,{\rm g\,cm^{-3}}}}

\def\mpl{{m_{\rm Pl}}}
\def\la{\mathrel{\mathpalette\fun <}}
\def\ga{\mathrel{\mathpalette\fun >}}
\def\fun#1#2{\lower3.6pt\vbox{\baselineskip0pt\lineskip.9pt
  \ialign{$\mathsurround=0pt#1\hfil##\hfil$\crcr#2\crcr\sim\crcr}}}

\begin{document}
\pagestyle{empty}
\begin{center}
\rightline{FERMILAB--Pub--92/176-A}
\rightline{August 1992}
\rightline{(submitted to {\it Physical Review Letters})}

\vspace{.2in}
{\bf INFLATION AT THE ELECTROWEAK SCALE} \\

\vspace{.2in}
Lloyd Knox$^{1}$ and Michael S. Turner$^{1,2,3}$ \\

$^1${\it Department of Physics,\\
The University of Chicago, Chicago, IL  60637-1433}\\

$^2${\it  Department of Astronomy \& Astrophysics,\\
Enrico Fermi Institute,
The University of Chicago, Chicago, IL 60637-1433}\\

$^3${\it NASA/Fermilab Astrophysics Center,
Fermi National Accelerator Laboratory, Batavia, IL  60510-0500}\\

\end{center}

\vspace{.3in}

\noindent{\bf ABSTRACT:}  We present a simple model
for slow-rollover inflation where the vacuum energy that drives inflation
is of the order of $G_F^{-2}$; unlike most models,
the conversion of vacuum energy to radiation
(``reheating'') is moderately efficient.  The scalar field responsible
for inflation is a standard-model singlet, develops a
vacuum expectation value of the order of
$4\times 10^6\GeV$, has
a mass of order $1\GeV$, and can play a role in electroweak
phenomena.  We also discuss models where the energy scale of
inflation is somewhat larger than the electroweak
scale, but still well below the unification scale.

\newpage
\pagestyle{plain}
\setcounter{page}{1}
\newpage

Over the past decade cosmologists have come to realize that
elementary-particle physics plays a very important role in cosmology:
Microphysical events that took place during the earliest moments of
the Universe ($t\ll 10^{-5}\sec$) and involved very high energies
($E\gg \GeV$) likely hold the key to
understanding some of the most puzzling
features of the Universe today.  For example,
baryon-number, $C$, and $CP$ violating interactions
occurring early on can explain the net baryon number of
the Universe (baryogenesis \cite{baryo}); the ubiquitous dark matter
may be comprised of relic elementary particles (particle
dark matter \cite{dark});
an early period of rapid expansion may account for
the smoothness and spatial flatness of the Universe
(inflation \cite{inflation});
and a variety of early Universe scenarios have
been proposed to explain the origin of the density inhomogeneities necessary
to seed the formation of structure in the Universe
(inflation, cosmic strings \cite{cs}, and textures \cite{textures}).

Until recently it appeared that the ``input
microphysics'' for these intriguing speculations
involved energies of the order of
$10^{14}\GeV$ or larger (grand-unification
scale), well beyond the ``reach'' of terrestrial
experiments.   However, scenarios for baryogenesis based upon physics
at the electroweak scale have been put forth \cite{ewbaryo},
and here we propose a simple model for inflation
at the electroweak scale.   The appeal of early Universe
scenarios based upon physics at the electroweak scale, of course,
is the possibility that the underlying physics can
be tested in the near future, e.g., at LEP (CERN), at the Tevatron
(Fermilab), or at the SSC.

Historically, inflation \cite{guth} developed from an attempt to
solve the monopole problem associated with grand-unified
theories (extreme overproduction of
magnetic monopoles during the GUT phase transition \cite{mp}),
and thus involved unification-scale energies.
Further, because the baryon asymmetry of the Universe
must be produced {\it after} inflation and most scenarios
for baryogenesis involve superheavy particles
and unification-scale physics, it seemed necessary that
inflation involve a very-high energy scale.  Indeed, in essentially
all models of inflation the vacuum energy that drives
inflation is of the order of $(10^{14}\GeV )^4$ \cite{models}.
Moreover,  in some models of inflation---chaotic
inflation \cite{chaotic}, inflation based upon
a simple supergravity model \cite{fla}, and extended inflation
\cite{extended}---the energy scale of inflation is set by
requiring density perturbations of an appropriate
size, and in these models that energy scale {\it must} be
of the order of the unification scale.

In this {\it Letter} we discuss a simple model of inflation
where the vacuum energy that drives inflation can be as small as
the electroweak scale ($\approx 1\TeV$).  We
begin the description of our model by reviewing the
requirements that a ``successful'' model of inflation
must satisfy \cite{inflation,KT}:

\noindent(1)  Sufficient inflation to
solve the horizon and flatness problems.  This
corresponds to $N \sim 30 + \ln (\trh /1\TeV )$ e-foldings
of the cosmic-scale factor during inflation,
where $\trh$ is the temperature at the beginning of
the post-inflation, radiation-dominated epoch.
When the energy scale of inflation is smaller,
the required amount of inflation is less.

\noindent(2)  Density perturbations of appropriate
size:  $\delta\rho /\rho \approx 10^{-5}$
(most difficult requirement to satisfy).
Density perturbations must be large enough to
initiate structure formation
and small enough to be consistent with the smoothness
of the cosmic background radiation (CBR).  Moreover, the
recent detection of temperature anisotropies
in the CBR on angular scales larger than about $10^\circ$
by the COBE DMR \cite{DMR} allows us to be more precise
about the amplitude of the density perturbations.

\noindent(3)  Sufficiently-high reheat temperature.
The Universe must be radiation dominated by
the epoch of primordial nucleosynthesis ($\trh
\gg 1\MeV$) so that nucleosynthesis proceeds in
the usual way, and, hot enough after inflation
for baryogenesis to take place, as
any pre-inflation baryon asymmetry is diluted
exponentially by the enormous entropy
release associated with reheating.
While it was thought that baryogenesis
required temperatures in excess of $10^{10}\GeV$ or so,
interesting models now exist where baryogenesis
occurs at the electroweak scale \cite{ewbaryo}
and temperatures as low as $1\GeV$ \cite{hall,raby}.

\noindent(4)  The abundance of unwanted, massive relics
such as monopoles, gravitinos, and oscillating
scalar fields produced after
inflation (e.g., during reheating) must be very small.  In order that
such nonrelativistic relics not contribute too much mass density
today, their energy density after inflation must be less than
$(10^{-8}\GeV /\trh )$ times that of radiation.
This is easier to satisfy when the
energy scale of inflation is lower:
Not only is the constraint less stringent, but many of the
dangerous relics are too heavy to be produced at such a
low energy scale.  Monopoles provide a good example:
In unification-scale models of inflation there is the
concern that GUT symmetry breaking occurs after inflation,
so that the monopole problem is not solved.

\noindent(5)  An integral part of a sensible particle-physics
model---or better yet, a testable part!

We denote the scalar field responsible for inflation
by $\phi$; as is well appreciated, in slow-rollover inflation
$\phi$ must be very weakly coupled in order to
satisfy the density-perturbation
constraint \cite{inflation}.  At the energy scale of interest, $\phi$ must
be a gauge singlet of the effective Lagrangian \cite{note}.
For simplicity, we take its scalar potential to
be of the Coleman-Weinberg type \cite{CW}, where
the symmetry-breaking minimum is generated by radiative
corrections,
\begin{equation} \label{cw}
V(\phi) = \frac{B\sigma^4}{2} + B\phi^4 \left[
        \ln (\phi^2 /\sigma^2) - \frac{1}{2}\right] ;
\end{equation}
other simple polynomial potentials can also be used
(e.g., $V = V_0 -\alpha\phi^4 + \beta\phi^5$ \cite{poly}).
Here $\sigma$ is the global minimum of the potential and
$B$ is a dimensionless coupling whose value must be
about $10^{-15}-10^{-14}$ to achieve density
perturbations of the appropriate size.
({\it All} models of inflation
have such a small coupling constant whose fundamental
understanding is still lacking.)
Coleman-Weinberg potentials are very flat near $\phi =0$,
$V \simeq {\cal M}^4 - b\phi^4$ where ${\cal M}^4 = B\sigma^4/2$, $b=
|\ln (\phi^2/\sigma^2)|\,B$, and for this reason
have been used often in models of inflation \cite{cwmodels}.

If, for the moment, we ignore the coupling of $\phi$ to other
fields, the equation of motion for $\phi$
in the expanding Universe is,
\begin{equation} \label{kg}
{\ddot\phi} + 3H\dot\phi + V^\prime = 0;
\end{equation}
where we have also assumed that the $\phi$ field is
homogeneous (or at least constant over a region
of space of order the Hubble radius).
During inflation $\phi$ ``rolls'' very slowly---but
inevitably---toward $\phi=\sigma$, and as
it does its potential energy dominates the
energy density of the Universe
driving a nearly constant expansion rate,
\begin{equation} \label{exprate}
H^2 = {8\pi V(\phi ) \over 3\mpl^2} \simeq {4\pi B\sigma^4\over 3\mpl^2};
\end{equation}
where $\mpl \equiv G^{-1/2} =1.22\times 10^{19}\GeV$ is the
Planck mass.  During the slow-roll phase, when $\phi$ is
near the origin ($\phi \la \phi_e$ \cite{end}),
the $\ddot\phi$ term can be neglected
so that $\dot\phi \simeq -V^\prime /3H$.  Using this approximation,
it follows that during the time it takes the scalar field to
evolve from $\phi$ to the minimum of its potential
the cosmic-scale factor grows by $N(\phi )$ e-foldings:
\begin{equation} \label{N}
N(\phi ) \simeq {8\pi\over \mpl^2}\int_\phi^{\phi_e}
        {V(\phi )d\phi \over -V^\prime} \simeq
        {\pi\over 2|\ln (\phi^2/\sigma^2)|}\,
        {\sigma^4\over \mpl^2\phi^2};
\end{equation}
where $|\ln (\phi^2/\sigma^2)|\approx 60$ is approximately
constant during the slow roll.
In order to achieve the 30 or so e-foldings of inflation
required the initial value of the scalar field must
be less than $\sigma^2/30\mpl \simeq 10^{-14}\sigma ({\cal M}/\TeV)$;
this is the least attractive feature of our model.

During the slow-roll phase density fluctuations
arise due to quantum fluctuations
in the scalar field $\phi$.  The amplitude of the perturbation
on a given scale $\lambda$, when that scale crosses inside the
horizon, is roughly \cite{inflation}
\begin{equation} \label{density}
\left( {\delta\rho\over \rho}\right)_{{\rm HOR},\lambda}
\sim \left({H^3\over V^\prime}\right)_{N_\lambda}
\sim \sqrt{B}N_\lambda^{3/2};
\end{equation}
where subscript $N_\lambda$ indicates that the
quantity is to be evaluated when the scale of interest crossed
outside the horizon during inflation, which occurs
$N_\lambda \simeq 21 + \ln (\trh /1\TeV) + \ln
(\lambda /\Mpc )$ e-foldings before the end of inflation.
To achieve $\delta\rho /\rho \approx 10^{-5}$, $B$ must be of order $10^{-15}$.

The quadrupole anisotropy in the CBR temperature
detected by the COBE DMR \cite{DMR}
allows us to be more precise about the value of $B$.
Expanding the CBR temperature on the sky
in spherical harmonics, the
quadrupole temperature anisotropy is
related to $a_2^2 \equiv \sum_m \langle a_{2m}^2\rangle$
(the average, over all observations positions
in the Universe, of the sum of the $l=2$
spherical-harmonic amplitudes squared) and
the inflationary potential:
\begin{equation} \label{quad}
\left( {\Delta T\over T}\right)_Q^2 = {a_2^2\over 4\pi} =
        {32\pi \over 45}{V^3\over {V^\prime}^2\mpl^6}\approx
        {2|\ln (\phi^2/\sigma^2)| B \over 45\pi^2}
        N_{\lambda}^3 .
\end{equation}
Setting $N_\lambda \sim 30$, the
scale of relevance for the quadrupole anisotropy,
and taking $(\Delta T/T)_Q \simeq 6\times 10^{-6}$,
we find that $B= 6\times 10^{-15}$.  This result is
relatively insensitive to the scale of inflation---for
$\sigma = 10^{16}\GeV$, $|\ln (\phi^2/\sigma^2)|
\sim 15$ and $N_\lambda
\sim 50$, which leads to $B\simeq 3\times 10^{-15}$---but
very sensitive to the value of $(\Delta T/T)_Q$, which
is probably uncertain by a factor of two.

One last remark about density perturbations; from
Eq. (\ref{density}) we see that the perturbations
are not quite scale invariant, $(\delta\rho /\rho )_{{\rm HOR}, \lambda}
\propto N_\lambda^{3/2}$.  Expanding $(\delta\rho /\rho)_{{\rm HOR},\lambda}$
about the mean of the galaxy scale ($1\Mpc$) and the present horizon
scale ($10^4\Mpc$) we find that
$(\delta\rho /\rho )_{{\rm HOR},\lambda} \propto \lambda^{0.06}$
(corresponding to a power spectrum $|\delta_k|^2\propto
k^n$ with $n=0.88$).  This has the effect of
depressing perturbations on small
scales relative to large scales by about a factor of
two, and may be important, as some
numerical simulations indicate that an exactly scale-invariant
spectrum of density perturbations normalized to the
COBE DMR quadrupole has too much power on small scales
\cite{gelb}.

Quantum fluctuations during inflation also give rise
to a spectrum of gravitational waves \cite{gw}; these gravitational
waves cross the horizon after inflation with an amplitude
of the order of $H/\mpl \sim 2\times 10^{-32}\,({\cal M}/\TeV)^2$,
orders of magnitude smaller than in models where the
scale of inflation is of the order of the
unification scale---and far too small to be detected.

Finally, consider reheating, the conversion of the
vacuum energy to thermal radiation.
After its slow roll, the $\phi$ field begins to oscillate
about the minimum of its potential,
and the vacuum energy that drives inflation
is converted into coherent scalar-field oscillations
(corresponding to a condensate
of nonrelativistic $\phi$ particles).  Reheating
takes place when the $\phi$ particles decay into
light fields, which, through their decays and interactions,
eventually produce a thermal bath of radiation.  During the
epoch of coherent $\phi$ oscillations the Universe
is matter dominated and the energy density trapped in the $\phi$
field decreases as the cube of the scale factor.  The reheat temperature
is determined by the decay time of the scalar
field oscillations, which is given by the
inverse of the decay width $\Gamma$ of the $\phi$ \cite{inflation}.
If $\Gamma \la H$, the coherent oscillation phase
is relatively long and the reheat temperature
$\trh \simeq \sqrt{\mpl\Gamma} < {\cal M}$,
corresponding to less than 100\% conversion of vacuum energy
to radiation.  Inefficient reheating is the rule for
slow-rollover inflation.  On the other hand, if $\Gamma \ga H$,
$\phi$ oscillations decay rapidly,
and $\trh \simeq {\cal M}$, corresponding to 100\%
conversion of vacuum energy to radiation.
Next we discuss why reheating is typically
very inefficient in slow-rollover inflation,
and how it becomes more efficient as the scale of inflation
is decreased.

Suppose the $\phi$ field couples to a light, Majorana fermion with
Yukawa coupling $g$; its decay width $\Gamma = g^2m_\phi /4\pi$,
where $m_\phi^2 = V^{\prime\prime}(\sigma )
= 8\sqrt{2B}{\cal M}^2 \simeq \GeV^2\,({\cal M}/\TeV)^2$.
The condition for efficient reheating is
\begin{equation} \label{rh}
{\Gamma\over H} = \sqrt{3g^4\over 8\pi^3}\,{\mpl\over \sigma}
\simeq \left( {g\over 2\times 10^{-6}} \right)^2 \,{\TeV\over {\cal M}} \ga 1.
\end{equation}
The condition for efficient
reheating depends upon the scale of inflation:
The larger the scale of inflation, the larger the
value of $g$ required for efficient reheating;
for ${\cal M} = 10^{14}\GeV$,
good reheating requires $g\ga 0.5$.

Next, consider the other constraints to the Yukawa coupling
$g$.   In order not to
spoil the flatness of $V(\phi )$, the radiative corrections
due to the fermion that couples to $\phi$ must be small:
This requires $g^4 \ll B$ or $g\ll 3\times 10^{-4}$.   Further, the
coupling to the $\phi$ field will give it a mass of
order $m_f\sim g\sigma$, which must be less than
half the mass of the $\phi$.  This provides
the stricter constraint, $g\la \sqrt{2B}$, and
illustrates how reheating and density perturbations
work at cross purposes:   Reheating is better
for a larger value of $B$, but density perturbations
require a very small value for $B$.

By saturating the bound $g\la \sqrt{2B} \sim 10^{-7}$,
we can express the maximum achievable reheat
temperature as a function of the scale of inflation:
\begin{equation}    \label{eff}
{\trh ({\rm max})\over {\cal M}} \sim
{\sqrt{\Gamma \mpl}\over {\cal M}} \simeq
B^{5/8} \,\sqrt{\mpl\over {\cal M}} \approx
0.1 \sqrt{{1\TeV \over {\cal M}}}.
\end{equation}
For ${\cal M}\sim \TeV$, $\trh ({\rm max})$
is of order $100\GeV$, and $\trh ({\rm max})$ grows
only as $\sqrt{\cal M}$:  for the canonical
scale of inflation, ${\cal M}\sim 10^{14}\GeV$,
$\trh ({\rm max})$ is only $3\times 10^7\GeV$.
Other modes of reheating are possible;
e.g., $\phi \rightarrow 2\chi$ ($\chi$ is
another scalar field), through interaction
terms of the form, ${\cal L}_{\rm int}
= \beta\phi\chi^2$ or $\lambda\phi^2\chi^2$.
Up to factors of order unity, the
same result obtains for the maximum achievable reheat
temperature, i.e., Eq. (\ref{eff}) \cite{bs,tmax}.

Let us squarely address the least attractive
feature of our model, the small initial value of $\phi$
required for sufficient inflation,
$\phi_i \la \sigma^2/30\mpl \simeq 10^{-14}\sigma\,({\cal M}/
\TeV )$.  Many models of slow-rollover inflation
require a small initial value for $\phi$; the very small
value required here traces to the very-low energy scale
of inflation:  For comparison, taking ${\cal M}\sim 10^{14}\GeV$,
$\phi_i \la 10^{-3}\sigma$.  This problem
can be mitigated by degrees by increasing the scale of inflation.

In order to achieve inflation in our model there must be
regions of the Universe where the value of the
$\phi$ field is very small; such regions will
undergo inflation.   In regions
where the value of the $\phi$ field is not
small, there will be no inflation.  After
inflation, the regions where $\phi$ was sufficiently small
have grown exponentially in size---and with
plausible assumptions about the distribution
of the initial value of $\phi$ the inflated regions
should occupy most of the physical volume of the Universe.

Such a small initial value for $\phi$ is not spoiled by
the quantum fluctuations in $\phi$, which are of the
order of $H/2\pi \sim 2\times 10^{-7}\sigma^2/\mpl \sim
10^{-19}\sigma$.  Thermal fluctuations will spoil
such localization:  $\langle \phi^2 \rangle_T^{1/2}
\sim T\simeq \TeV \sim 10^{-4}\sigma$.  However,
it can be argued that $\phi$
is so weakly coupled it is not in thermal contact with
the Universe; indeed, this
argument has been used for other models of inflation \cite{fla}.

Another way of insuring that the small initial value
of the $\phi$ field is not spoiled by thermal fluctuations
is to arrange that inflation begin ``cold,''
$T\ll 1\TeV$.  There are plausible ways
that this might occur.   If the Universe, or a small
portion of it, were so negatively curved that
it became curvature dominated early on, say
at a temperature $T_{\rm CD}$, then the temperature when inflation
begins is $T_{\rm inflate}\sim \TeV^2/T_{\rm CD}$, which can easily be
small enough to render the thermal fluctuations impotent.
(Within the spirit
of ``generic'' initial conditions, one would expect
the curvature radius at the Planck epoch to be of the
order of the Hubble radius, in which case $T_{\rm CD}\sim
\mpl$ and $T_{\rm inflate}\sim 10^{-13}\GeV$.)
Or, the Universe can become
matter dominated long before inflation, e.g., by monopoles
produced at the GUT phase transition, or other massive
relics produced copiously in the early
Universe.  And of course, it is not necessary that
the Universe have any radiation in it prior to
inflation:  It could have begun cold.

Finally, we comment briefly on the phenomenology of our model.
Because $\phi$ is very weakly coupled, it must
be an $SU(3)_C\otimes SU(2)_L\otimes U(1)_Y$ singlet; however,
it can {\it indirectly} influence electroweak physics.
The vacuum expectation value (VEV) of $\phi$,
$\sigma \sim 4\times 10^6\GeV$, can
induce a negative mass-squared for the Higgs field (call it $\psi$)
that does lead to electroweak symmetry breaking, through
a coupling $\lambda \psi^2\phi^2$ \cite{trick}.  A negative mass-squared
of order $1\TeV$ requires $\lambda \sim 10^{-7}$.  Since
the radiative corrections to the $V(\phi )$ due to $\psi$
are of order $\lambda^2/4\pi^2 \sim 10^{-15}\sim B$, they
are about the right size to account for the $\phi$
field's symmetry-breaking potential.  The VEV of $\phi$ can give
rise to particle masses, e.g., righthanded neutrinos; in this
case reheating can take place by $\phi$ decays into
righthanded neutrinos and their subsequent decays into light leptons.
If the scale of inflation is raised slightly,
${\cal M} \sim 200\TeV$, the mass of the $\phi$
particle is of order several hundred $\GeV$.
In this case, reheating can take place through
$\phi$ decays into electroweak Higgs and their subsequent
decays into the particles of the standard model.

In sum, we have presented a simple model
of slow-rollover inflation where the vacuum energy that
drives inflation can be as small as the electroweak
scale---orders of magnitude smaller than in
previous models.  Inflation at a
low-energy scale has a number of attractive features:
reheating is more efficient;
the monopole problem is more easily solved; and last, but
not least, such a model is potentially testable.

\vskip 1.5cm
\noindent  We thank J. Harvey, B. Mertens, S. Raby, J. Rosner, and M. Worah
for useful discussions.  This work was supported in part by the
DOE (at Chicago and Fermilab) and by the NASA through
NAGW-2381 (at Fermilab).


\begin{thebibliography}  {inflation}
\bibitem{note}  If $\phi$ were coupled
to a gauge field with strength $g$, then radiation
corrections due to gauge boson loops would lead
to $B\sim g^4$ and $B\gg 10^{-14}$ unless
$g\la 10^{-4}$.
\bibitem{poly}  The potential $V(\phi ) = V_0 -\alpha \phi^4
+\beta \phi^5$ has its global, zero-energy minimum at $\sigma =
(5V_0/\alpha )^{1/4}$, where $\beta = 4\alpha /5\sigma$.
A model of inflation similar to the one presented here
can be constructed for $\alpha = 3\times 10^{-13}$.
Since the $\phi^5$ term is nonrenormalizable, we imagine that
it arises in an effective Lagrangian due to physics at a
much higher energy scale.
\bibitem{end}  The end of the slow-roll epoch occurs when
$|V^{\prime\prime}| \approx 9H^2$; for our model,
$\phi_e/\sigma \simeq \sqrt{\pi/|\ln (\phi^2/\sigma^2)}(\sigma /\mpl )
\simeq 10^{-13}$.
\bibitem{tmax} There are possibilities for improving the reheating
efficiency.  The strict constraint to the coupling of the
field into which the $\phi$ decays, based upon its mass
being less than half that of the $\phi$, can be circumvented
if there are two contributions its mass,
of opposite sign, so that its coupling to the $\phi$
can exceed this bound.  Further,
a subtlety of reheating could be useful:
The highest temperature reached after inflation occurs just
when the $\phi$ particles begin to decay, long before
most of them decay, $T_{\rm max} \simeq \sqrt{\trh {\cal M}}$.
As the temperature of the Universe falls from $T_{\rm max}$ to
$\trh$ the entropy per comoving volume increases
as $T^{-5}$; the reheat temperature is
defined to be the temperature when essentially all
the $\phi$ particles have decayed and the entropy per
comoving volume becomes constant \cite{inflation}.  In spite of the
entropy production it may be possible to
take advantage of this higher temperature.
\bibitem{trick} In this regard we have followed the models of Shafi and
Vilenkin \cite{models} and Pi \cite{models} where
the VEV of the $\phi$ was used to induce GUT symmetry breaking.
\end{thebibliography}
\end{document}